\documentclass[twocolumn,prb,aps,showpacs]{revtex4}
\usepackage{amsmath}
\usepackage{graphicx}

\begin{document}
\title{Skyrmion crystals in the pseudo-spin-1/2 Bose-Einstein condensates}
\author{Cong Zhang}
\author{Wenan Guo}
\author{Shiping Feng}
\author{Shi-Jie Yang\footnote{Corresponding author: yangshijie@tsinghua.org.cn}}
\affiliation{Department of Physics, Beijing Normal University, Beijing 100875, China}
\begin{abstract}
Exact two-dimensional solutions are constructed for the pseudo-spin-1/2 Bose-Einstein condensates which are described by the coupled nonlinear Gross-Pitaevskii equations where the intraspecies and interspecies coupling constants are assumed to be equal. The equations are decoupled by means of re-combinations of the nonlinear terms of the hyperfine states according to the spatial dimensions. These stationary solutions form various spin textures which are identified as skyrmion crystals. In a special case, the crystal of skyrmion-antiskyrmion pairs is formed in the soliton limit.
\end{abstract}
 \maketitle

\section{introduction}
A skyrmion is a particle-like topologically nontrivial soliton \cite{Choi} which is studied in a variety of research fields, especially in the condensed matter physics such as the quantum Hall effects \cite{Doucot,Schmeller,Moon,Sondhi,Timm}, the liquid crystals \cite{Wright}, and the helical ferromagnets \cite{Neubauer,Yu,Zang}. It is noted early that skyrmions in three dimensions (3D) have physical properties of baryons while skyrmions in two-dimensional (2D) play an important role in condensed matter systems. In recent experiments, it has been demonstrated that the 2D skyrmion spontaneously appears as the ground state in the helical magnets\cite{Choi} and in the quantum Hall systems at fillings slightly away from unity.

Since the experimental realization of quantized vortices in alkali atomic Bose-Einstein condensates (BECs), phenomena related to the internal degrees of freedom are appealing for the investigation of topological objects\cite{Yang,Ramachandhran,Huhtam-1,Ruben,Huhtam-2,Scherer,Yi-1,Kasamatsu,Kawakami}. Furthermore, precise spin manipulation techniques have been developed to prepare topological spin structures of interest and provide unique opportunities to study their stability and dynamics \cite{Bretin,Schweikhard,Leslie}. The skyrmion-skyrmion interaction may lead to the formation of a Skyrme crystal. Meanwhile, the periodic skyrmion crystal configurations could be used to model nuclear matter \cite{Skyrme,Klebanov,Bogdanov}. Such a state was recently observed in a neutron scattering experiment\cite{Muhlbauer}.

In this paper we attempt to seek exact stationary solutions to the 2D coupled nonlinear Gross-Pitaevskii equations (GPEs) which describe the pseudo-spin-1/2 BECs. The intra- and inter-species interactions are assumed to be equal and the system is exposed to a uniform external field. We propose a method to decouple the GPEs by re-combinations of the nonlinearity and the spatial dimensions. Periodic solutions are obtained which are shown to form the skyrmion crystals in the pseudo-spin representation. In Sec.II we describe the method. The main results are displayed in Sec.III. Section IV contains a brief summary.

\section{method}
We consider the 2D pseudo-spin-1/2 BEC in a uniform external potential ($V({\bf r})=0$) by assuming the intra- and inter-species interaction strength $\gamma_{11}=\gamma_{12}\equiv\gamma$ and the two-component atoms have the same mass $m_1=m_2\equiv m$. The Hamiltonian then has the "pseudospin" $SU(2)$ symmetry\cite{Kasamatsu}. By adopting the units of $\hbar=m=1$, the stationary GPEs for the mean-field order parameter $\Psi=(\psi_1,\psi_2)^\textrm{T}$ are
\begin{equation}
  \mu_j\psi_j({\bf r})=[-\frac{1}{2}\nabla ^2 +\gamma(|\psi_1({\bf r})|^2+|\psi_2({\bf r})|^2)] \psi_j({\bf r}),\label{stationary}
\end{equation}
where $\mu_j$ ($j=1,2$) denote the chemical potentials. In order to solve the coupled Eqs.(\ref{stationary}), we decompose the wave functions according to the dimensions as
\begin{equation}
\left\{
      \begin{array}{c}
\psi_1(x,y)=X_1(x)Y(y)+ i X(x)Y_1(y)\\
\psi_2(x,y)=X_2(x)Y(y)+i X(x)Y_2(y)
      \end{array}\right.,
      \label{normal}
\end{equation}
where $X_1(x),X_2(x),X(x)$ and $Y_1(y),Y_2(y),Y(y)$ are real functions of a single variable. By imposing the restrictions of \begin{eqnarray}
\left\{\begin{array}{c}
|X_1(x)|^2+|X_2(x)|^2=B_1^2\\
|Y_1(y)|^2+|Y_2(y)|^2=B_2^2
      \end{array}\right.,
\end{eqnarray}
where $B_1$ and $B_2$ are real constants, the total density can be seperated as
\begin{equation}
 \rho({\bf r})=B_1^2X^2(x)+B_2^2Y^2(y).
\end{equation}
Substituting Eq.(\ref{normal}) into Eq.(\ref{stationary}) and equaling the real and the imaginary parts, respectively, we obtain a set of ordinary differential equations for $X(x)$ and $Y(x)$ which can be solved self-consistently by making use of the unique properties of the Jacobian elliptical functions\cite{Abramowitz}. The details are described in the next section.

According to the relations of three non-singular Jacobian elliptical functions, $\textrm{sn},\textrm{cn}$, and $\textrm{dn}$, we construct three forms of solutions to the Eqs.(\ref{stationary}). They exhibit periodic crystals of the skyrmions or skyrmion-anti-skyrmion pairs, respectively. The periods of the wavefunctions are $k=4\cdot j K(m)$, where $K(m)$ is the complete elliptic integral of the first kind with modulus $0<m<1$ and $j$ is a integer which denotes the number of periods.

\section{stationary solutions}
\subsection{Type A}
We first consider the following form of solution,
\begin{equation}
\left\{\begin{array}{c}
\psi_1(x,y)=A \textrm{sn}(kx,m) \textrm{cn}(ky,m)+i B\textrm{cn}(kx,m) \textrm{sn}(ky,m)\\
\psi_2(x,y)=D \textrm{cn}(kx,m) \textrm{cn}(ky,m).\label{sccsAcc}
      \end{array}\right.
\end{equation}
This wavefunction automatically satisfies the periodic boundary conditions in the $x$- and $y$-direction due to the properties of the Jacobian elliptical functions. By substituting Eq.(\ref{sccsAcc}) into the stationary Eq.(\ref{stationary}) and using the identities between the Jacobian elliptic functions, we obtain the relations:
\begin{equation}
A^2=B^2=\frac{1}{2}D^2,
\end{equation}
and
\begin{eqnarray}
\left\{\begin{array}{c}
\mu_1=\frac{1}{2} k^2(2-3m^2)\\
\mu_2=k^2(1-2m^2),
      \end{array}\right.\label{mu1A}
\end{eqnarray}
\begin{eqnarray}
\gamma=-\frac{k^2 m^2}{A^2}.
\end{eqnarray}
From $\gamma<0$, we note that this solution applies to the BEC with attractive interactions. The difference between the chemical potentials $\mu_i$ ($i=1,2$) of the two components can be induced by a Zeeman energy. The total number of atoms is related to the amplitude of the wavefunction by
\begin{eqnarray}
N=\frac{2 A^2}{m^2}[m^2-1+\frac{E(m)}{K(m)}],\label{N1A}
\end{eqnarray}
where $E(m)$ is the complete elliptic integral of the second kind.

\begin{figure}[htbp]
\begin{center}
\includegraphics*[width=9cm]{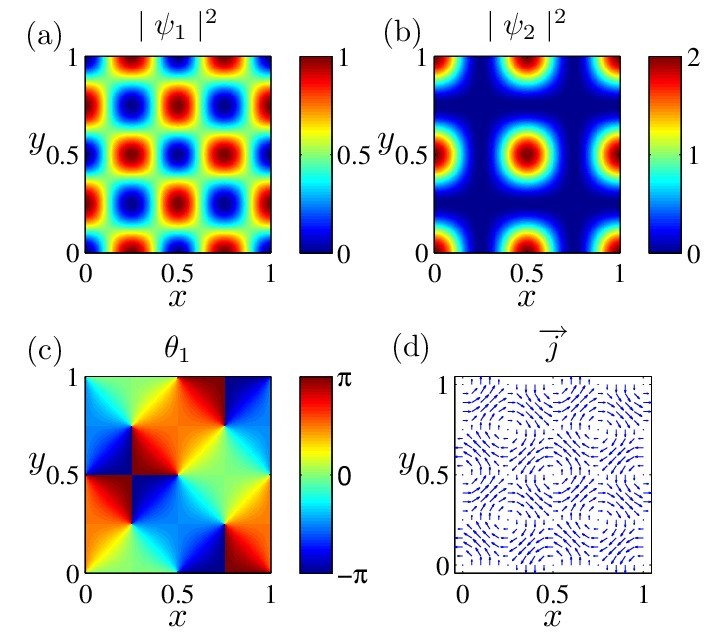}
\caption{(Color online) The density profiles of the $\psi_1$ (a) and $\psi_2$ (b) hyperfine states in the state (\ref{sccsAcc}). The parameters are $A=1$, $\gamma=-11.3670$, $\mu_1=28.4175$, $\mu_2=22.7340$ and $m=0.5$. The two-species have different particle number and different chemical potential. (c) The phase distribution of the complex $\psi_1$ component. (d) The total particle current density. The number of period of the wavefunction $j=1$.}
\end{center}
\end{figure}

Figure 1(a) and (b) display the checkerboard distributions of density of each hyperfine state for $A=1$, $\gamma=-11.3670$, $\mu_1=28.4175$, $\mu_2=22.7340$ and $m=0.5$. The phase of $\psi_1$ shown in Fig.1(c) reveals a periodic vortex-anti-vortex pair structure. The phase difference implies a particle current density which is defined by\cite{Kasamatsu,Ueda}
\begin{eqnarray}
\textbf{j}=\frac{1}{2i} \sum_{j=1,2} [\psi_{j}^*(\boldsymbol{\nabla} \psi_{j})-(\boldsymbol{\nabla} \psi_{j}^*)\psi_{j}].
\end{eqnarray}
Fig.1(d) displays the vortex-anti-vortex structure of the total particle current.

In order to explore the topological structure of the state, we examine the normalized spinor $\chi({\bf r})$ which is defined by $\Psi({\bf r})=\sqrt{\rho({\bf r})}\chi({\bf r})$. The spin vector $\textbf{S}$ for the condensates are $\textbf{S}(\bf r)=\chi^\dag\boldsymbol{\sigma}\chi$, with $\boldsymbol{\sigma}$ the Pauli matrices. Fig.2(a) displays the spin texture which forms a skyrmion crystal. The topological charge or the Pontryagian index of the skyrmion is an invariant\cite{Yang},
\begin{eqnarray}
Q= \int_{\textrm{cell}} q(\textbf{r})dxdy,\label{Q}
\end{eqnarray}
where the topological charge density
\begin{eqnarray}
q(\textbf{r})=\frac{1}{4 \pi}\textbf{S} \cdot (\partial_x \textbf{S} \times \partial_y \textbf{S}).
\end{eqnarray}
Fig.2(b) shows distribution of $q(\textbf{r})$ which exhibits a periodic structure. We note that in one period of the state (\ref{sccsAcc}), there are four skyrmions. Each unit cell has a total topological charge $Q=1$. By the way, the anti-skyrmion crystal can be obtained by simply take the complex conjugacy of the state (\ref{sccsAcc}).

\begin{figure}[htp]
\begin{center}
\includegraphics*[width=9cm]{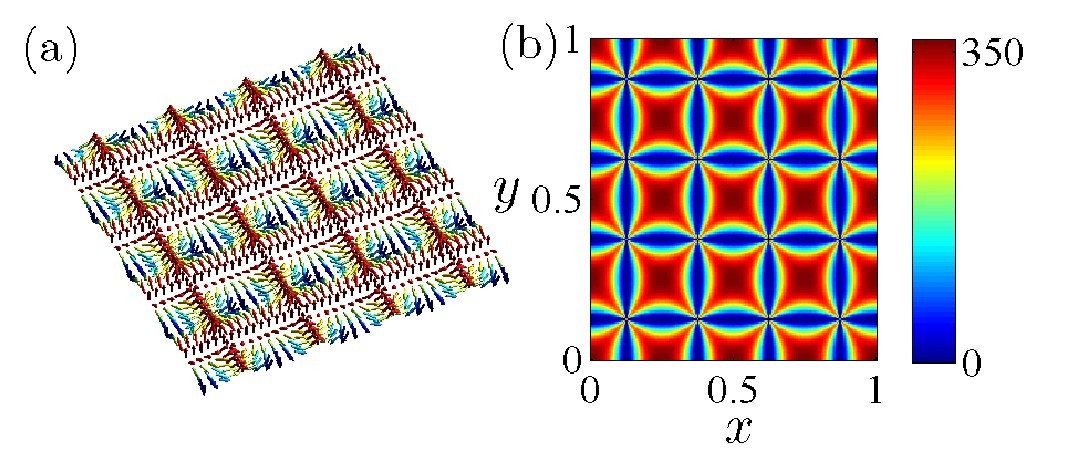}
\caption{(Color online) (a) The spin texture ${\bf S}({\bf r})$ of the state (\ref{sccsAcc}) forms a skyrmion crystal. The number of period $j=2$. (b) The distribution of the topological charge density $q(\textbf{r})$. The total topological charge of each cell is unity.}
\end{center}
\end{figure}

\subsection{Type B}
Next we consider the solution for $\gamma>0$ which is fulfilled by the following configuration,
\begin{equation}
\left\{\begin{array}{c}
\psi_1(x,y)=A \textrm{sn}(kx,m) \textrm{cn}(ky,m)+i B\textrm{cn}(kx,m)\textrm{sn}(ky,m)\\
\psi_2(x,y)=D \textrm{sn}(kx,m)\textrm{sn}(ky,m).\label{sccsBss}
      \end{array}\right.
\end{equation}
We obtain the relations:
\begin{equation}
A^2=B^2=\frac{1}{2}D^2,
\end{equation}
and
\begin{eqnarray}
\left\{\begin{array}{c}
\mu_1=\frac{1}{2} k^2(2+m^2)\\
\mu_2=k^2(1+m^2)
      \end{array}\right.,
\end{eqnarray}
\begin{eqnarray}
\gamma=\frac{k^2 m^2}{A^2}.
\end{eqnarray}
The total number of atoms is related to the amplitude of the wavefunction by
\begin{eqnarray}
N=\frac{2 A^2}{m^2}[1-\frac{E(m)}{K(m)}].
\end{eqnarray}

\begin{figure}[t]
\begin{center}
\includegraphics*[width=9cm]{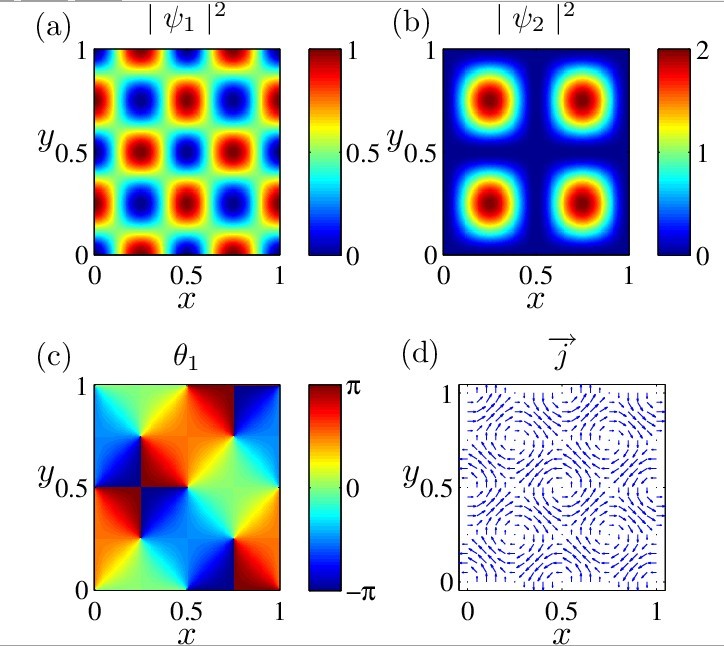}
\caption{(Color online) The same as in Fig.1 for the state (\ref{sccsBss}). The parameters are $A=1$, $\gamma=11.3670$, $\mu_1=56.8351$, $\mu_2=51.1516$ and $m=0.5$.}
\end{center}
\end{figure}

Figure 3 display the results as those as in Fig.1 for $j=1$. The parameters are $A=1$, $\gamma=11.3670$, $\mu_1=56.8351$, $\mu_2=51.1516$ and $m=0.5$. The spin texture forms a periodic crystal of anti-skyrmions as shown in Fig.4(a). Our calculation shows that each anti-skyrmion has a total topological charge $Q=-1$ (Fig.4(b)).
\begin{figure}[htp]
\begin{center}
\includegraphics*[width=9cm]{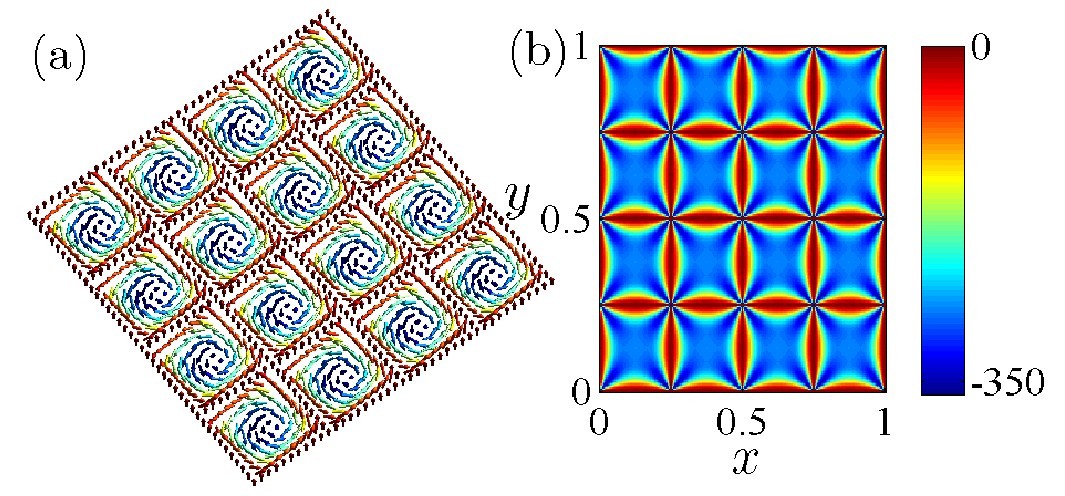}
\caption{(Color online) The same as in Fig.2 for the repulsive case of the state (\ref{sccsBss}). The total topological charge in each cell is $Q=-1$.}
\end{center}
\end{figure}

\subsection{Type C}
We finally show an example of solution that is generally not a skyrmion as $m<1$ while tends to be a skyrmion in the soliton limit ($m\rightarrow 1$). The solution is of the form
\begin{equation}
\left\{\begin{array}{c}
\psi_1(x,y)=A \textrm{sn}(kx,m) \textrm{dn}(ky,m)+i B\textrm{dn}(kx,m) \textrm{sn}(ky,m)\\
\psi_2(x,y)=D \textrm{dn}(kx,m) \textrm{dn}(ky,m).\label{sddsCdd}
      \end{array}\right.
\end{equation}
We obtain the relations:
\begin{equation}
A^2=B^2=\frac{1}{2}m^2 D^2,
\end{equation}
and
\begin{eqnarray}
\left\{\begin{array}{c}
\mu_1=\frac{1}{2} k^2(2m^2-3)\\
\mu_2=k^2(m^2-2),
      \end{array}\right.
\end{eqnarray}
\begin{eqnarray}
\gamma=-\frac{k^2}{A^2}.
\end{eqnarray}
Obviously, the solution applies to the attractive BECs. The total number of atoms is related to the amplitude of the wavefunction by
\begin{eqnarray}
N=2 A^2 \frac{E(m)}{K(m)}.
\end{eqnarray}

\begin{figure}[htbp]
\begin{center}
\includegraphics*[width=9cm]{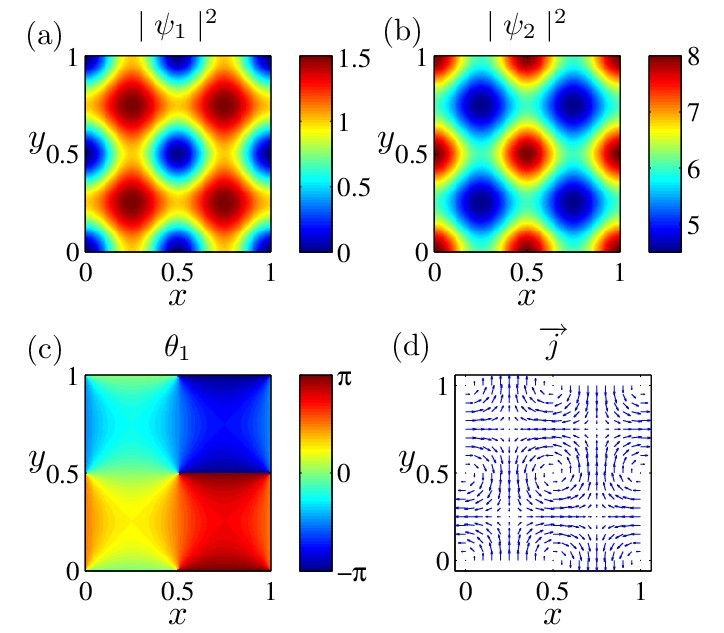}
\caption{(Color online) The same as in Fig.1 for the state (\ref{sddsCdd}). The parameters are $A=1$, $\gamma=-45.4681$, $\mu_1=-56.8351$, $\mu_2=-79.5691$ and $m=0.5$.}
\end{center}
\end{figure}

\begin{figure}[tp]
\begin{center}
\includegraphics*[width=9cm]{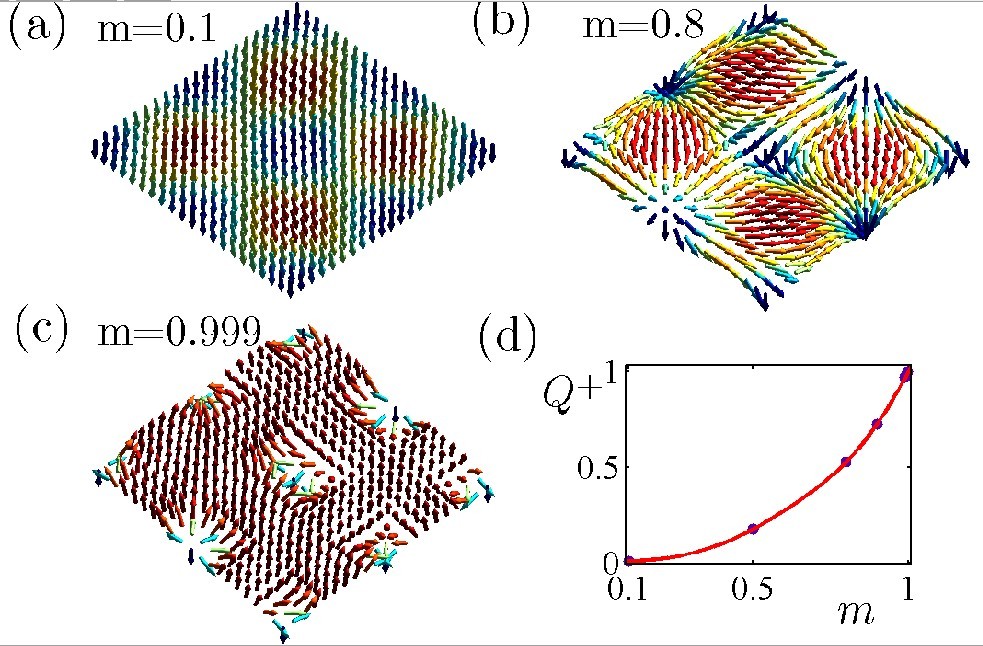}
\caption{(Color online) (a)-(c) The spin texture of the state (\ref{sddsCdd}) for $m=0.1,0.8$, and the solitonic limit $m=0.999$. (d) the total charge defined by (\ref{Q}) versus the modulus $m$. As $m\rightarrow 1$ the spin texture becomes a skyrmion-anti-skyrmion crystal.}
\end{center}
\end{figure}

\begin{figure}[htbp]
\begin{center}
\includegraphics*[width=9cm]{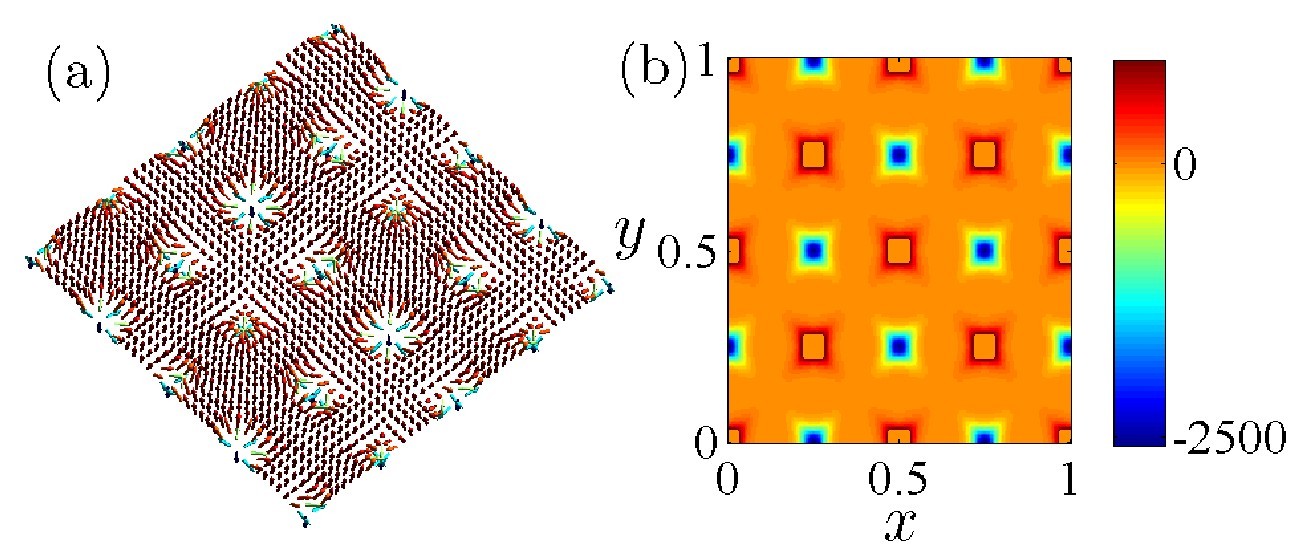}
\caption{(Color online) The same as in Fig.2 for the state (\ref{sddsCdd}) in the limit of $m\rightarrow 1$. The total topological charge in each cell $Q=\pm 1$ which forms a dipole crystal.}
\end{center}
\end{figure}

Figure 5 display the results as those in Fig.1 for the state (\ref{sddsCdd}). The parameters are chosen as $A=1$, $\gamma=-45.4681$, $\mu_1=-56.8351$, $\mu_2=-79.5691$ and $m=0.5$. Despite of the similarity, the state (\ref{sddsCdd}) is essentially distinct to former two states. To clarify this point, we show in Fig.6 the spin textures for various modulus $m$. We find that the total topological charge defined by (\ref{Q}) for each domain is not an integer. In a period, the spin texture does not form a skyrmion since the spin vector ${\bf S({\bf r})}$ does not rotate a complete circle around the $z$-axis, as shown in Fig.6(a) and (b). Nevertheless, in the solitonic limit we find the spin texture still form a skyrmion-anti-skyrmion crystal in comparison to the former solutions. Fig.6(d) reveals that the total topological charge in each period is alternatively $Q=\pm 1$ as $m\rightarrow 1$. This configuration is shown in Fig.7, where the charge density forms a dipole crystal.

\section{Summary}
In summary, we have presented a class of exact solutions to the coupled GPEs which describe the pseudo-spin-1/2 BECs. The spin textures exhibit periodic crystals consisting of skyrmions ($Q=1$ or $Q=-1$) or skyrmion-anti-skyrmion pairs.

This work is supported by the funds from the Ministry of Science and Technology of China under Grant
No.2012CB821403. W. G. is supported by NSFC under grant No.11175018.

\end{document}